\documentclass[12pt]{article}
\title{On Charged Fields with Group Symmetry\\ and Degeneracies of Verlinde's Matrix $S$}
\author{Michael M\"uger\thanks{Partially supported by the Studienstiftung des deutschen Volkes and by CEE.} \\ Dipartimento di Matematica, Universit\`{a} di Roma ``Tor Vergata''\\ Via della Ricerca Scientifica, 00133 Roma, Italy\\ Email: mueger@axp.mat.uniroma2.it}

\newlength{\dinwidth}
\newlength{\dinmargin}
\usepackage{amssymb}

\def\1#1{{\bf #1}}
\def\2#1{{\cal #1}}
\def\3#1{{\sl #1}}
\def\4#1{{\tt #1}}
\def\5#1{{\sf #1}}
\def\6#1{{\mathfrak #1}}
\def\7#1{{\mathbb #1}}

\newcommand{\be}{\begin{equation}}
\newcommand{\ee}{\end{equation}}
\newcommand{\ba}{\begin{array}}
\newcommand{\ea}{\end{array}}
\newcommand{\bea}{\begin{eqnarray}}
\newcommand{\eea}{\end{eqnarray}}
\newcommand{\bean}{\begin{eqnarray*}}
\newcommand{\eean}{\end{eqnarray*}}
\newcommand{\nn}{\nonumber}

\newcommand{\ve}{\varepsilon}

\newcommand{\impl}{\Rightarrow}

\newcommand{\restr}{\upharpoonright}
\newcommand{\ol}{\overline}

\newcommand{\del}{\partial}

\newcommand{\qed}{\ \ $\blacksquare$}
\newcommand{\qft}{quantum field theory}
\newcommand{\qfts}{quantum field theories}
\newcommand{\poinc}{Poincar\'{e}}

\newcommand{\npb}{Nucl. Phys. \1B}
\newcommand{\cmp}{Commun. Math. Phys. }

\newcommand{\rmp}{Rev. Math. Phys. }
\newcommand{\jfa}{J. Funct. Anal. }

\newtheorem{defin}{Definition}[section]
\newtheorem{lemma}[defin]{Lemma}
\newtheorem{prop}[defin]{Proposition}
\newtheorem{theorem}[defin]{Theorem}
\newtheorem{coro}[defin]{Corollary}
\newtheorem{conj}[defin]{Conjecture}

\newcommand{\bdefin}{\begin{defin}}
\newcommand{\blemma}{\begin{lemma}}
\newcommand{\bprop}{\begin{prop}}
\newcommand{\btheor}{\begin{theorem}}
\newcommand{\bcoro}{\begin{coro}}
\newcommand{\edefin}{\end{defin}}
\newcommand{\elemma}{\end{lemma}}
\newcommand{\eprop}{\end{prop}}
\newcommand{\etheor}{\end{theorem}}
\newcommand{\ecoro}{\end{coro}}
\newcommand{\bconj}{\begin{conj}}
\newcommand{\econj}{\end{conj}}

\newcommand{\prf}{{\it Proof. }}
\newcommand{\rem}{{\it Remark. }}
\newcommand{\rems}{{\it Remarks. }}

\newcommand{\sectreset}[1]{\section{#1}\setcounter{equation}{0}}

\begin{document}
\maketitle\noindent

\abstract{We consider the complete normal field net with compact symmetry group
constructed by Doplicher and Roberts starting from a net of local observables in 
$\ge 2+1$ spacetime dimensions and its set of localized (DHR) representations. We prove
that the field net does not possess nontrivial DHR sectors, provided the observables 
have only finitely many sectors. Whereas the superselection structure
in $1+1$ dimensions typically does not arise from a group, the DR construction 
is applicable to `degenerate sectors', the existence of which (in the rational case) is 
equivalent to non-invertibility of Verlinde's S-matrix. We prove Rehren's conjecture that
the enlarged theory is non-degenerate, which implies that every degenerate theory is
an `orbifold' theory. Thus, the symmetry of a generic model `factorizes'
into a group part and a pure quantum part which still must be clarified.}

\sectreset{Introduction}
A few years ago a long-standing problem in local quantum physics \cite{haag} (algebraic
\qft) was solved in \cite{dr2}, where the conjecture \cite{dhr1,dr0} was proved that the 
superselection structure of the local observables can always be described in terms of a 
compact group. This group (gauge group of the first kind) acts by automorphisms on a 
net of field algebras which generate the charged sectors from the vacuum and obey normal
Bose and Fermi commutation relations. From the mathematical point of view this amounts to
a new duality theory for compact groups \cite{dr6} which considerably improves on the 
old Tannaka-Krein theory. These results rely on a remarkable chain of arguments 
\cite{dr4,dr5,dr3,dr1} which we cannot review here. We refer to the first two sections
of \cite{dr2} for a relatively non-technical overview of the construction and restrict
ourselves to a short introduction to the problem in order to set the stage for our
considerations.

Our starting point is a net of local observables, i.e.\ an inclusion preserving
map $\2O\mapsto\2A(\2O)$ which assigns to each double cone $\2O$ (the set of these is
denoted by $\2K$) in spacetime the algebra of observables measurable in $\2O$. More 
specifically, identifying the abstract local algebras with their images in a faithful 
vacuum representation $\pi_0$, we assume the $\2A(\2O)$ to be von Neumann algebras 
acting on the Hilbert space $\2H_0$. The $C^*$-algebra $\2A$ generated by all $\2A(\2O)$
is called the quasilocal algebra. As usual the property of Einstein causality (if 
$\2O_1,\2O_2$ are mutually spacelike double cones then $\2A(\2O_1)$ and $\2A(\2O_2)$ 
commute elementwise) is strengthend by requiring Haag duality
\be \2A(\2O)'=\2A(\2O')^- \ \ \forall\2O\in\2K ,\ee
where $\2A(\2O')$ is the $C^*$-algebra generated by 
$\2A(\tilde{\2O}),\, \2K\ni\tilde{\2O}\subset\2O'$. Typically one requires
\poinc\ or conformal covariance but these properties will play no essential role for our
considerations, apart from their being used to derive the Property B (cf.\ Subsect.\
\ref{bose} below) which is needed for the analysis of the superselection structure.

We restrict our attention to superselection sectors which are localizable in arbitrary
double cones, i.e.\ representations $\pi$ of the quasilocal algebra $\2A$ satisfying the
DHR criterion \cite{dhr3,dhr4}:
\be \pi\restr\2A(\2O')\cong\pi_0\restr\2A(\2O') \ \ \forall\2O\in\2K .\label{dhr}\ee
These representations are called locally generated since they are indistinguishable from
the vacuum when restricted to the spacelike complement of a double cone.
Given a representation of this type, Haag duality implies \cite{dhr3} for any double 
cone $\2O$ the existence of a unital endomorphism of $\2A$ which is localized in $\2O$ 
(in the sense that $\rho(A)=A\ \forall A\in\2A(\2O')$) such that 
$\pi\cong\pi_0\circ\rho\equiv\rho$. This is an important fact since endomorphisms can
be composed, thereby defining a composition rule for this class of representations.
Whereas (non-surjective) endomorphisms are not invertible, there are left inverses, i.e.\
(completely) positive linear maps $\phi_\rho: \2A\rightarrow\2A$ such that 
$\phi_\rho(\rho(A)B\rho(C))=A\phi_\rho(B)C$, in particular $\phi_\rho\circ\rho=id$. 
Localized endomorphisms obtained from DHR representations are
transportable, i.e.\ given $\rho\in\Delta$ there is an equivalent morphism localized in 
$\2O$ for every $\2O\in\2K$. Furthermore, given two localized endomorphisms, one can 
construct operators $\ve(\rho_1,\rho_2)$ which intertwine $\rho_1\rho_2$ and 
$\rho_2\rho_1$ and thereby formalize the notion of particle interchange (whence the name
statistics operators). For $\rho$ an irreducible morphism, 
$\phi_\rho(\ve(\rho,\rho))=\lambda_\rho\11$ gives rise via polar decomposition 
$\lambda_\rho=\omega_\rho/d_\rho$ to a phase and a positive number. From 
here on the analysis depends crucially on the number of 
spacetime dimensions. In $\ge 2+1$ dimensions \cite{dhr3,dhr4} the statistics operators 
$\ve(\rho_1,\rho_2)$ are uniquely defined and satisfy $\ve(\rho,\rho)^2=\11$ such that
one obtains, for each morphism $\rho$, a unitary representation of the permutation
group in $\2A$ via $\sigma_i\mapsto\rho^{i-1}(\ve(\rho,\rho))$. Furthermore, the 
statistics phase and dimension satisfy $\omega_\rho=\pm 1$ and 
$d_\rho\in\7N\cup\{\infty\}$. The statistics phase $\omega_\rho$ distinguishes 
representations with bosonic and fermionic character, and the statistical dimension 
$d(\rho)$ measures the degree of parastatistics. Ignoring morphisms with infinite 
dimension, which are considered pathological, we denote by $\Delta$ the semigroup of all
transportable localized morphisms with finite statistics. 

The analysis which was sketched above was motivated by the preliminary investigations
conducted in \cite{dhr1}. There the starting point is a net of field algebras 
$\2O\mapsto\2F(\2O)$ acted upon by a compact group $G$ of inner symmetries (gauge group
of the first kind):
\be \alpha_g(\2F(\2O))=\2F(\2O) .\ee
The field algebra acts irreducibly on a vacuum Hilbert space $\2H$ and the gauge
group is unbroken, i.e.\ represented by unitary operators $U(g)$ in a strongly continuous
way: $\alpha_g(F)=Ad\, U(g)(F)$. (Compactness of $G$ need in fact not be postulated, as 
it follows by \cite[Thm. 3.1]{dl} if the field net satisfies the split property.)

The field net is supposed to fulfill Bose-Fermi commutation relations, i.e.\ any local
operator decomposes into a bosonic and a fermionic part $F=F_+ + F_-$ such that
for spacelike separated $F$ and $G$ we have
\be [F_+,G_+]=[F_+,G_-]=[F_-,G_+]=\{F_-,G_-\}=0 .\label{i-commrel}\ee
The above decomposition is achieved by
\be F_\pm = \frac{1}{2}(F\pm\alpha_k(F)) ,\label{i-fpm}\ee
where $k$ is an element of order 2 in the center of the group $G$. $V\equiv U_k$ is the 
unitary operator which acts trivially on the space of bosonic vectors and like $-\11$ on
the fermionic ones. To formulate this locality requirement in a way more convenient for 
later purposes we introduce the twist operation $F^t=ZFZ^*$ where
\be Z=\frac{1+iV}{1+i}, \ \ \ (\Rightarrow Z^2=V)\ee
which leads to $ZF_+Z^*=F_+,\ ZF_-Z^*=iVF_-$ implying $[F,G^t]=0$. 
The (twisted) locality postulate (\ref{i-commrel}) can now be stated simply as
\be \2F(\2O)^t\subset\2F(\2O')'   .\label{i-commrel2}\ee
In analogy to the bosonic case, this can be strengthened to twisted duality:
\be \2F(\2O)^t = \2F(\2O')'   \label{i-twduality} .\ee
The observables are now defined as the fixpoints under the action of $G$:
\be \2A(\2O) = \2F(\2O)^G = \2F(\2O)\cap U(G)' .\label{avono}\ee
The Hilbert space $\2H$ decomposes as follows:
\be \2H=\bigoplus_{\xi\in\hat{G}} \2H_\xi \otimes \7C^{d_\xi} ,\label{decomp}\ee
where $\xi$ runs through the equivalence classes of finite dimensional continuous unitary
representations of $G$ and $d_\xi$ is the dimension of $\xi$. 
The observables and the group $G$ act reducibly according to
\be\ba{cccccc} A &=& \bigoplus_{\xi\in\hat{G}} & \pi_\xi(A) & \otimes & \11 ,\\
   U(g) &=& \bigoplus_{\xi\in\hat{G}} & \11_{\2H_\xi} & \otimes & U_\xi(g) ,\ea
\label{decomp2}\ee
where $\pi_\xi$ and $U_\xi$ are irreducible representations of $\2A$ and $G$, 
respectively. As a consequence of twisted duality for the fields, the restriction of
the observables $\2A$ to a simple sector (subspace $\2H_\xi$ with $d_\xi=1$), in 
particular the vacuum sector,
satisfies Haag duality. Since the unitary representation of the \poinc\ group commutes
with $G$, the restriction of $\2A$ to $\2H_0$ satisfies all requirements for a net of
observables in the vacuum representation in the above sense.
As shown in \cite{dhr1}, the irreducible representations of $\2A$
in the charged sectors are globally inequivalent but strongly locally equivalent to 
each other (i.e.\ $\pi_1\restr\2A(\2O')\cong\pi_2\restr\2A(\2O')$),
in particular they satisfy the DHR criterion. Obviously it is not necessarily true that 
the decomposition (\ref{decomp2}) contains all equivalence classes of DHR representations
(take
$\2F=\2A,\ \2H=\2H_0,\ G=\{e\}$). This completeness is true, however, if the field net 
$\2F$ has trivial representation theory (equivalently `quasitrivial $1$-cohomology'),
see \cite{r4}. It was conjectured in \cite{dr0} that every net of observables arises 
as a fixpoint net such that the representation of $\2A$ on $\2H$ contains all sectors, 
which furthermore means that the tensor category of DHR sectors with finite statistics 
is isomorphic to the representation category of a compact group $G$.
Under the restriction that all transportable localized morphisms are automorphisms,
which is equivalent to $G$ being abelian, this was proved in \cite{dhr2}. After
the early works \cite{dr0,r-cpgd} the final proof in complete generality 
\cite{dr4,dr5,dr3,dr1,dr6,dr2} turned out to be quite difficult, which is 
perhaps not too surprising in view of the nontriviality of the result.

In the next section we prove a few complementary results concerning the DR-construc\-tion
in $\ge 2+1$ dimensions. It is natural to conjecture that the DR field net does not 
possess localized superselection sectors provided it is complete, i.e.\ contains charged
fields generating all DHR sectors (with finite statistics) of the observables. 
Whereas, at first sight, this may 
appear to be an obvious consequence of the uniqueness result \cite{dr2} for the complete
normal field net we have unfortunately been able to give a proof only for the case of 
a finite gauge group, i.e.\ for rational theories. Under the same assumption we show that
the complete field net can also be obtained by applying the DR construction to an 
intermediate, i.e.\ incomplete field net. Whereas in higher dimensions the restriction to
finite gauge groups is quite unsatisfactory, our results have a useful application
to the low dimensional case to which we now turn.

In $1+1$ dimensions there are in particular two interesting classes of models. The first
consists of purely massive models, many of these being integrable. Concerning these it
has been shown recently \cite{mue3} that they do not have DHR sectors at all as long
as one insists on the assumption of Haag duality. As to conformally covariant models, 
which constitute the other class of interest, the situation is quite different in that
it has been shown \cite{bmt} that 
positive-energy representations are necessarily of the DHR type due to local normality 
and compactness of the spacetime. It is particularly this class which we have in mind
in our $2d$ considerations, but the conformal covariance will play no role.
Whereas in $\ge 2+1$ dimensions one has $\ve(\rho_2,\rho_1)^*=\ve(\rho_1,\rho_2)$, in
$1+1$ dimensions these statistics operators are a priori different intertwiners 
between $\rho_1\rho_2$ and $\rho_2\rho_1$. This phenomenon accounts for the occurrence 
of braid group statistics and provides the motivation for defining the 
{\it monodromy operators}:
\be \ve_M(\rho_1,\rho_2)=\ve(\rho_1,\rho_2)\,\ve(\rho_2,\rho_1) ,\ee
which measure the deviation from permutation group statistics.
An irreducible morphism $\rho$ is said to be degenerate if $\ve_M(\rho,\sigma)=\11$ for
all $\sigma$. Given two irreducible morphisms $\rho_i,\rho_j$ one obtains the 
$\7C$-number valued {\it statistics character} \cite{khr2} via
\be Y_{ij}\11=d_i d_j\, \phi_j(\ve_M(\rho_i,\rho_j)^*) .\ee
(Here $\phi_j$ is the left inverse of $\rho_j$ and the factor $d_i d_j$ has been 
introduced for later convenience.) The numbers $Y_{ij}$ depend only on the sectors,
such that the matrix $(Y_{ij})$ can be considered as indexed by the set of equivalence
classes of irreducible sectors. The matrix $Y$ satisfies the following identities:
\be Y_{0i}=Y_{i0}=d_i ,\ee
\be Y_{ij}=Y_{ji}=Y_{i\bar{\jmath}}^*=Y_{\bar{\imath}\bar{\jmath}}^* ,\ee
\be Y_{ij}=\sum_k N_{ij}^k \frac{\omega_i\omega_j}{\omega_k} d_k ,\ee
\be \frac{1}{d_j}Y_{ij}Y_{kj}=\sum_m N_{ik}^m Y_{mj} .\ee
Here $[\rho_{\bar{\imath}}]$ is the conjugate morphism of $[\rho_i]$ and
$N_{ij}^k\in\7N_0$ is the multiplicity of $[\rho_k]$ in the decomposition of
$[\rho_i\rho_j]$ into irreducible morphisms. The matrix of statistics characters is of
particular interest if the theory is {\it rational}, i.e.\ has only a finite number of 
inequivalent irreducible representations. Then, as proved by Rehren \cite{khr2}, the 
matrix $Y$ is invertible iff there is no degenerate morphism besides the trivial one
which corresponds to the vacuum representation. In the non-degenerate case the number
$\sigma=\sum_i d_i^2\omega_i^{-1}$ satisfies $|\sigma|^2=\sum_i d_i^2$ and the matrices
\be S=|\sigma|^{-1} Y,\ \ \ T=\left(\frac{\sigma}{|\sigma|}\right)^{1/3} Diag(\omega_i)
\ee
are unitary and satisfy the relations
\be S^2=(ST)^3=C, \ \ TC=CT=T ,\label{sl2z}\ee
where $C_{ij}=\delta_{i,\bar{\jmath}}$ is the charge conjugation matrix. That is,
$S$ and $T$ constitute a representation of the modular group $SL(2,\7Z)$.
Furthermore, the `fusion coefficients' $N_{ij}^k$ are given by the Verlinde relation
\be N_{ij}^k=\sum_m \frac{S_{im}S_{jm}S_{km}^*}{S_{0m}} .\label{i-v}\ee
As was emphasized in \cite{khr2}, these relations hold independently of conformal
covariance in every (non-degenerate) two dimensional theory with finitely many DHR
sectors. This
is remarkable, since the equation (\ref{i-v}) first appeared \cite{v} in the context of 
conformal \qft\ on the torus, where the $S$-matrix by definition has the additional 
property of describing the behavior of the conformal characters 
$tr_{\pi_i}e^{-\tau L_0}$ under the inversion $\tau\rightarrow -1/\tau$. 

The equations (\ref{sl2z}, \ref{i-v}) do not hold if the matrix $Y$ is not
invertible, i.e.\ when there are degenerate sectors. One can show that the set of
degenerate sectors is stable under composition and reduction into irreducibles
(Lemma \ref{trivm}).
It thus constitutes a closed subcategory of the category of DHR endomorphisms to
which one can apply the DR construction of charged fields. In Sect.\ 4 we will prove
Rehren's conjecture in \cite{khr2} that the resulting `field' net has no degenerate
sectors. Furthermore, we will prove that the enlarged theory is rational, provided 
that the original one is. These results imply that the above Verlinde-type analysis is
in fact applicable to $\2F$.

\sectreset{On the Reconstruction of Fields from Observables}
Our first aim in this section will be to prove the intuitively reasonable fact that a 
complete field net associated (in $\ge 2+1$ dimensions) with a net of observables does 
not possess localized superselection sectors. For technical reasons we have been able to
give a proof only for rational theories. This result, which may not be too useful in 
itself, will after some preparations be the basis of our proof of a conjecture by Rehren
(Thm.\ \ref{no-deg2}). Furthermore, we show that the construction of the complete field 
net `can be done in steps', that is, one also obtains the complete field net by applying
the DR construction to an intermediate, thus incomplete, field net and its DHR sectors. 
For the sake of simplicity we defer the treatment of the general case for a while and 
begin with the purely bosonic case.

\subsection{Absence of DHR Sectors of the Complete Field Net: Bose Case}\label{bose}
The superselection theory of a net of observables is called purely bosonic if all
DHR sectors have statistics phase $+1$. In this case the charged fields which generate
these sectors from the vacuum are local and the fields associated with different sectors 
can be chosen to be relatively local. Then the Doplicher-Roberts construction \cite{dr2}
gives rise to a local field net $\2F$, which in addition satisfies Haag duality. Thus it
makes sense to consider the DHR sectors of $\2F$ and to apply the DR construction to
these. (In analogy to \cite{dhr3,dhr4} one requires $\2F$ to satisfy the technical 
`property B' \cite{dhr3}, which can be derived
\cite{da} from standard assumptions, in particular positive energy. Since a
DR field net is \poinc\ covariant with positive energy \cite[Sect. 6]{dr2}, provided this
is true for the vacuum sector and the DHR representations of the observables, we may 
take the property B for granted also for $\2F$.) 

We cite the following definitions from \cite{dr2}:
\bdefin Given a net $\2A$ of observables and a vacuum representation $\pi_0$, a normal 
field system with gauge symmetry, $\{\pi,\2F,G\}$, consists of a representation $\pi$ of
$\2A$ on a Hilbert space $\2H$ containing $\pi_0$ as a subrepresentation on 
$\2H_0\subset\2H$, a compact group $G$ of unitaries on $\2H$ leaving $\2H_0$ pointwise
fixed and a net $\2O\mapsto\2F(\2O)\subset\2B(\2H)$ of von Neumann algebras such that
\begin{description}
\item[$\alpha$)] the $g\in G$ induce automorphisms $\alpha_g$ of $\2F(\2O),\ \2O\in\2K$ 
with $\pi(\2A(\2O))$ as fixed-point algebra,
\item[$\beta$)] the field net $\2F$ is irreducible,
\item[$\gamma$)] $\2H_0$ is cyclic for $\2F(\2O)\ \forall\2O\in\2K$,
\item[$\delta$)] there is an element $k$ in the center of $G$ with $k^2=e$ such that the
net $\2F$ obeys graded local commutativity for the $\7Z_2$-grading defined by $k$, cf.\ 
(\ref{i-commrel}, \ref{i-fpm}).
\end{description}
\label{nfn}\edefin

\bdefin A field system with gauge symmetry $\{\pi,\2F,G\}$ is complete if each
equivalence class of irreducible representations of $\2A$ satisfying (\ref{dhr})
and having finite statistics is realized as a subrepresentation of $\pi$, i.e.\ $\pi$
describes all relevant superselection sectors. \edefin

For a given net of observables $\2A$ we denote by $\Delta$ the set of all transportable
localized morphisms with finite statistics. Let $\Gamma$ be a closed semigroup of 
localized {\it bosonic} endomorphisms and let $\2F$ be the associated local field net. 
Now let $\Sigma$ be a closed semigroup of localized endomorphisms of $\2F$. After
iterating the DR construction again we are faced with the following situation.
There are three nets $\2A,\,\2F,\,\tilde{\2F}$ acting faithfully and
irreducibly on the Hilbert spaces $\2H_0\subset\2H\subset\tilde{\2H}$, respectively, such
that Haag duality holds (twisted duality in the case of $\tilde{\2F}$). 
The nets $\tilde{\2F}$ and $\2F$ are normal field nets with respect to the nets $\2F$ and
$\2A$, respectively, in the sense of Def.\ \ref{nfn}. Thus there are
representations $\pi$ of $\2A$ on $\2H$ and $\tilde{\pi}$ of $\2F$ on 
$\tilde{\2H}$, respectively, such that 
$\tilde{\pi}\circ\pi(\2A)\subset\tilde{\pi}(\2F)\subset\tilde{\2F}$. Furthermore, 
there are strongly compact groups $G$ and $\tilde{G}$ of unitaries on $\2H$ and 
$\tilde{\2H}$, respectively, acting as local symmetries on $\2F$ and 
$\tilde{\2F}$, respectively, such that $\2F(\2O)^G=\pi(\2A(\2O)), \2O\in\2K$ and 
$\tilde{\2F}(\2O)^{\tilde{G}}=\tilde{\pi}(\2F(\2O)), \2O\in\2K$.
The following result is crucial:

\bprop Let the theory $\2A$ be rational (equivalently, let $G$ be finite). Then
the net $\tilde{\2F}$ is a normal field net w.r.t. the observables $\2A$. In 
particular, there is a strongly compact group $\ol{G}$ of unitaries on $\tilde{\2H}$ 
containing $\tilde{G}$ as a closed normal subgroup. $\ol{G}$ implements local 
symmetries of $\tilde{\2F}$ such that
$\tilde{\2F}(\2O)^{\ol{G}}=\tilde{\pi}\circ\pi(\2A(\2O))$. \label{forts1}\eprop
\prf Let $\ol{G}$ be the group of unitaries on $\tilde{\2H}$ implementing local 
symmetries of $\tilde{\2F}$ which leave $\2A$ pointwise and the algebras
$\2F(\2O), \2O\in\2K$ globally stable. Clearly, $\ol{G}$ is strongly closed and 
contains $\tilde{G}$ as a closed normal subgroup. We can now apply Prop.\ 3.1 of 
\cite{bdlr} to the effect that every element of $G$ extends to a unitarily
implemented local symmetry of $\tilde{\2F}$, thus an element of $\ol{G}$, such that
there is a short exact sequence
\be \11\rightarrow\tilde{G}\rightarrow\ol{G}\rightarrow G\rightarrow\11 .\ee
By assumption, $\tilde{G}$ is known to be compact in the strong topology which, of 
course, coincides with the topology induced from $\ol{G}$. The group $G$ being finite
it is clearly compact w.r.t.\ any topology. Compactness of $\tilde{G}$ and $G$ 
implies compactness of $\ol{G}$ (cf., e.g.,\ \cite[Thm.\ 5.25]{hr}).

It remains to prove the requirements $\beta)-\delta)$ of Def.\ \ref{nfn}. Now,
$\beta)$ and $\delta)$ are automatically true by \cite[Thm. 3.5]{dr2}. Finally, 
$\gamma)$, viz. the cyclicity of $\2H_0$ for $\tilde{\2F}(\2O),\,\2O\in\2K$ is also easy:
in application to $\2H_0$, $\tilde{\pi}(\2F(\2O))\subset\tilde{\2F}(\2O)$ gives a dense
subset of $\2H$, the image of which under the action of the charged (w.r.t. $\2F$) 
fields in $\tilde{\2F}$ is dense in $\tilde{\2H}$. \qed\\
\rem As to the general case of infinite $G$ we note that, $\tilde{G}$ being compact,
$\ol{G}$ is (locally) compact
iff $G=\ol{G}/\tilde{G}$ is (locally) compact in the quotient topology.
It is easy to show that the identical map from $G$ with the quotient topology to $G$ 
with the strong topology induced from $\2B(\2H)$ is continuous. Since we know that $G$ 
is compact w.r.t.\ to the latter and since both topologies are Hausdorff,
$G$ is compact w.r.t.\ to the former (and thus $\ol{G}$ is compact) iff the identical 
map is open. This would follow from an open mapping theorem \cite[Thm.\ 5.29]{hr} if
we could prove that the $G$ is locally compact and second countable with the quotient
topology. Clearly this idea can work only if the observables have at most countably 
many sectors. We hope to return to this problem in another paper.

We are now prepared to prove the absence of DHR sectors of the field net. Let 
$\Gamma=\Delta$, the set of all transportable localized morphisms of $\2F$
with finite statistics. Using Prop.\ \ref{forts1} we easily prove the following:
\btheor The complete (local) field net $\2F$ associated with a purely bosonic
rational theory has no DHR sectors with finite statistics. \label{nosect}\etheor
\prf Assuming the converse, the above proposition gives us a field net $\tilde{\2F}$ on 
a larger Hilbert space $\tilde{\2H}$, which obviously is also complete, since the
representation $\pi$ of $\2A$ on $\2H$ is a subrepresentation of $\tilde{\pi}\circ\pi$.
Thus, by \cite[Thm. 3.5]{dr2} both field systems are equivalent, that is, there is a
unitary operator $W: \2H\to\tilde{\2H}$ such that 
$W\pi(A)=\tilde{\pi}\circ\pi(A)W\ \forall A\in\2A$ etc. In view of the decomposition 
\be \pi=\bigoplus_{\xi\in\hat{G}} d_\xi \, \pi_\xi ,\ee
where the irreducible representations $\pi_\xi$ are mutually inequivalent, and similarly
for $\tilde{\pi}\circ\pi$, $\pi$ and $\tilde{\pi}$ can be unitarily equivalent only if 
$G=\tilde{G}$ and thus $\2F=\tilde{\2F}$. \qed\\
\rem After this paper was essentially completed I learned that this result (with the 
same restriction to finite groups) has been obtained about two years ago by R. Conti 
\cite{conti}.

We have thus, in the purely bosonic case, reached our first goal. Before we turn to the 
general situation we show that the construction of the complete field net `can be
done in steps', that is, one also obtains the complete field net by applying the DR
construction to an intermediate field net and its DHR sectors, again assuming that the
intermediate net is local (this is not required for the complete field net). 

\subsection{Stepwise Construction of the Complete Field Net: Bose Case}
The following lemma is more or less obvious and is stated here since it does not appear
explicitly in \cite{dr1,dr2}.
\blemma Let $\Gamma_1,\Gamma_2$ be subsemigroups of $\Delta$ which are both closed 
under direct sums, subobjects and conjugates and let $\2F_i,\ i=1,2$ be the 
associated normal field nets on the Hilbert spaces $\2H_i$ with symmetry groups
$G_i$ and $\pi_i$ the representations of $\2A$. If 
$\Gamma_1\subset\Gamma_2$ then there is an isometry $V: \2H_1\rightarrow\2H_2$ such that
\be V \pi_1(A)=\pi_2(A) V,\ \ A\in\2A ,\label{v1}\ee
\be VG_1 V^*=G_2E, \label{v2}\ee
\be V\2F_1 V^*=(\2F_2\cap\{E\}')E, \label{v3}\ee
where $E=VV^*$. Furthermore, there is a closed normal subgroup $N$ of $G_2$ such that
$E$ is the projection onto the subspace of $N$-invariant vectors in $\2H_2$ and 
$\{\pi_1,G_1,\2F_1\}$ is equivalent to $\{\pi_2^N,G_2/N,\2F_2^N\}$. \elemma
\prf As usual, the field theory $\2F_2$ is constructed by applying \cite[Cor.\ 6.]{dr2}
to the quadruple $(\2A,\Delta_2,\ve,\pi_0)$ and by defining $\2F(\2O)$ to be the
von Neumann algebra on $\2H_2$ generated by the Hilbert spaces 
$H_\rho,\ \rho\in\Delta_2(\2O)$. Let $E$ be the projection $[\2B_1\2H_0]$ where $\2B_1$
is the $C^*$-algebra generated by $H_\rho,\ \rho\in\Delta_1$. Trivially, $\2B_1$
maps $E\2H_2$ into itself. $\2B_1$ is stable under $G_2$ as each of the Hilbert spaces
$H_\rho$ is. This implies that $G_2$ leaves $E\2H_2$ stable. Restricting $\2B_1$ and 
$G_2$ to $E\2H_2$ one obtains the system 
$(E\2H_2,E\pi_2(\cdot)E,EU_2E,\rho\in\Delta_1\to EH_\rho E)$
which satisfies a) to g) of \cite[6.2]{dr2}. With the exception of g) all of these are
trivially obtained as restrictions. Property g) follows by appealing to 
\cite[Lemma 2.4]{dr3}. We can thus conclude from the uniqueness result of 
\cite[Cor.\ 6.2]{dr2}
that $(E\2H_2,E\pi_2E,EU_2E,\rho\in\Delta_1\to H_\rho)$ is equivalent to the system 
$(\2H_1,\pi_1,U_1,\rho\in\Delta_1\to H_\rho)$ obtained from the quadruple 
$(\2A,\Delta_2,\ve,\pi_0)$, that is, there is a unitary $V$ from $\2H_1$ to $E\2H_2$
such that $V\pi_1(A)=\pi_2(A)V, \ V\2F_1=\2B_1 V,\ VU_1=U_2V$.
Interpreting $V$ as an isometry mapping $\2H_1$ into $\2H_2$ we have
(\ref{v1}-\ref{v3}). The rest follows from \cite[Prop. 3.17]{dr2}. \qed

\blemma Let $\2O\mapsto\2F(\2O)$ be the field net associated to a subsemigroup 
$\Gamma$ of $\Delta$, closed under direct sums, subobjects and conjugates. Then every 
localized endomorphism $\eta\in\Delta$ of $\2A$ extends to an endomorphism 
$\tilde{\eta}$ of $\2F$ commuting with the action of the gauge group. 
If $\eta$ is localized in $\2O$ the same holds for $\tilde{\eta}$. \label{forts2}\elemma
\rem This result is of interest only if $\eta\not\in\Gamma$. Otherwise we already
know that $\eta$ extends to an inner endomorphism of $\2F$ by definition of the field 
algebra. \\
\prf By the preceding result we know that the field net $\2F=\2F_\Gamma$ is equivalent to
a subnet of the complete field net $\ol{\2F}=\2F_\Delta$. We identify $\2F$ with 
this subnet. By construction every localized endomorphism $\eta\in\Delta(\2O)$ of 
$\2A$ extends to an inner endomorphism of $\ol{\2F}$. i.e.\ there is a multiplet
of isometries $\psi_i\in\ol{\2F}(\2O),\ i=1,\ldots,d$
satisfying $\sum_i\psi_i\psi_i^*=\11,\ \psi_i^*\psi_j=\delta_{i,j}\11$ such that
$\hat{\eta}\circ\pi(A)=\pi\circ\eta(A)$ where
\be \hat{\eta}(\cdot) = \sum_i \psi_i \cdot \psi_i^* .\ee
Since $\hat{\eta}$ commutes with the action of $G$, it is easy to verify that 
$\hat{\eta}$ leaves $\2F=\ol{\2F}^N$ stable and thus restricts to an endomorphism of 
$\2F$ which extends $\eta$. This extension is not necessarily local, for 
$\hat{\eta}(F)=-F$ if $F$ is a fermionic operator localized spacelike to $\2O$ and 
$\eta$ is a fermionic endomorphism. This defect is easily remedied by defining
\be \tilde{\eta}=\left\{ \ba{cl} \hat{\eta}  & \mbox{if }\omega(\eta)=1 \\
    Ad\,V\circ\hat{\eta}   & \mbox{if }\omega(\eta)=-1 \ea \right. \ee
Clearly, $\tilde{\eta}$ has the desired localization properties and coincides with
$\eta$ on $\2A$. Transportability of $\tilde{\eta}$ is automatic as $W\in(\eta,\eta')$
implies $\pi(W)\in(\tilde{\eta},\tilde{\eta}')$. Finally the statistical dimensions
of $\eta$ and $\tilde{\eta}$ coincide as is seen using, e.g., the arguments in 
\cite{lo2}. \qed\\
\rem The preceding lemmas do not depend on the restriction to bosonic families $\Gamma$
of endomorphisms or on the finiteness of the gauge group.

\blemma Let $\2A$ be rational, let $\Gamma$ be a semigroup of bosonic endomorphisms and 
let $\2F$ be the associated (incomplete) local field net. Let $\Sigma$ be the semigroup 
of all localized endomorphisms of $\2F$. Then the associated DR-field net $\tilde{\2F}$
is a complete field net with respect to $\2A$. \label{c-compl}\elemma
\prf Let $\eta$ be a localized endomorphism of $\2A$. By the preceding lemma, there
is an extension (typically reducible) to a localized endomorphism $\tilde{\eta}$ of 
$\tilde{\2F}$. By Prop.\ \ref{forts1}, $\tilde{\2F}$ is a normal field net for $\2A$.
By completeness of $\tilde{\2F}$ with respect to endomorphisms of
$\2F$, $\tilde{\eta}$ is implemented by a Hilbert space in $\tilde{\2F}$
and there is a subspace $\2H_{\tilde{\eta}}$ of $\tilde{\2H}$ such that
$\tilde{\pi}\restr\2H_{\tilde{\eta}}\cong\tilde{\eta}$ as 
a representation of $\2F$. Restricting to $\2A$ and choosing an irreducible subspace 
$\2H_\eta$ we have $\pi_\Sigma\restr\2H_{\tilde{\eta}}\cong\pi_0\circ\eta$.
Thus $\tilde{\2F}$ is a complete field net for $\2A$. \qed

\btheor Let $\2A$ be a rational net of observables and let $\Gamma$ be a bosonic 
subsemigroup of $\Delta$ with the associated field net $\2F_\Gamma$.
Then the complete normal field net $\2F_{\Gamma,\Sigma}$ 
obtained from the net $\2F_\Gamma$ and its semigroup $\Sigma$ of all localized 
endomorphisms is equivalent to the complete normal field net $\2F_\Delta$. 
In particular the group $\ol{G}$ obtained in Lemma \ref{forts1} is isomorphic to the 
group belonging to $\2F_\Delta$. \label{isom}\etheor
\prf By Lemmas \ref{forts1} and \ref{c-compl}, $\2F_{\Gamma,\Sigma}$ is a complete normal
field net for $\2A$. The same trivially holding for $\2F_\Delta$, we are done since two 
such nets are isomorphic by \cite[Thm. 3.5]{dr2}.\qed

\subsection{General Case, Including Fermions}\label{22}
In the attempt to prove generalizations of Thm.\ \ref{nosect} for theories possessing 
fermionic sectors and of Thm.\ \ref{isom} for fermionic intermediate nets $\2F$ 
we are faced with the problem that it is not entirely obvious what these generalizations
should be. We would like to show the representation theory of a complete normal field 
net, which is now assumed to comprise Fermi fields, to be trivial in some sense. 
It is not clear a priori that the methods used in the purely bosonic case will lead to
more than, at best, a partial solution. Yet we will adopt a conservative strategy and 
try to adapt the DHR/DR theory to $\7Z_2$-graded nets. The fermionic version of 
Thm.\ \ref{isom} will vindicate this approach.

Clearly, the criterion (\ref{dhr}) makes sense also for $\7Z_2$-graded nets. Since 
things are complicated by the spacelike anticommutativity of fermionic operators,
the assumption of twisted duality for $\2F$ is, however, not sufficient to deduce that 
representations satisfying (\ref{dhr}) are equivalent to (equivalence classes) of
transportable endomorphisms of $\2F$. To make this clear, assume $\pi$ satisfies
(\ref{dhr}), and let $X^\2O: \2H_0\to\2H_\pi$ be such that 
$X^\2OA=\pi(A)X^\2O\ \ \forall A\in\2F(\2O')$. We would like to show that
$\rho(A)\equiv X^{\2O*}\pi(A)X^\2O$ maps $\2F(\2O_1)$ into itself if $\2O_1\supset\2O$.
Now, let $x\in\2F(\2O_1),\ y\in\2F(\2O_1')^t$, which implies $xy=yx$. We would like to
apply $\rho$ on both sides and use $\rho(y)=y$ to conclude that
$\rho(\2F(\2O_1))\subset{\2F(\2O_1')^t}'=\2F(\2O_1)$. As it stands, this argument 
does not work, since $\pi$ and thus $\rho$ are defined only on the quasilocal 
algebra $\2F$, but not on the operators $VF_-\in\2F^t$ which result from the 
twisting operation. Assume, for a moment, that the representation $\rho$ lifts
to an endomorphism $\hat{\rho}$ of the $C^*$-algebra $\hat{\2F}$ on $\2H$ generated by
$\2F$ and the unitary $V$, such that $\hat{\rho}(V)=V$ or, alternatively, 
$\hat{\rho}(V)=-V$. Using triviality of $\rho$ in restriction to $\2F(\2O_1')$ we then
obtain $\rho(\2F(\2O_1')^t)=\2F(\2O_1')^t$, which justifies the above argument. Now, in 
order for $\hat{\rho}(V)=\pm V$ to be consistent, we must have 
\be \rho\circ\alpha_k(A)=\rho(VAV)=\hat{\rho}(V)\rho(A)\hat{\rho}(V)=V\rho(A)V=
  \alpha_k\circ\rho(A) ,\ee
i.e.\ $\rho\circ\alpha_k=\alpha_k\circ\rho$. In view of 
$\rho(A)=X^{\2O*}\pi(A)X^\2O$ we can now claim:
\blemma There is a one-to-one correspondence between equivalence classes of:
\begin{description}
\item[a)] Representations of $\2F$ which are, for every $\2O\in\2K$, unitarily
equivalent to a representation $\rho$ on $\2H_0$ such that $\rho\restr\2F(\2O')=id$
and $\rho\circ\alpha_k=\alpha_k\circ\rho$ (where 
$Aut\,\2B(\2H_0)\ni\alpha_k\equiv Ad\,V$);
\item[b)] Transportable localized endomorphisms of $\2F$ commuting with $\alpha_k$.
\end{description}
\label{c-endos}\elemma
\rem In a) covariance of $\pi$ with respect to $\alpha_k$ is not enough. We need the 
fact that, upon transferring the representation to the vacuum Hilbert space via 
$\rho(A)=X^{\2O*}\pi(A)X^\2O,\ \alpha_k$ is implemented by the grading operator $V$.\\
\prf The direction b)$\impl$a) is trivial. As to the converse, by the above all that
remains to prove is extendibility of $\rho$ to $\hat{\rho}$. By the arguments in 
\cite[p. 121]{sum-nps} the $C^*$-crossed product (covariance algebra) 
$\2F\rtimes_{\alpha_k}\7Z_2$ is simple such that the actions of $\2F$ and $\7Z_2$ on
$\2H_0$ and $\2H_\pi$ via $\pi_0=id,\ V$ and $\pi,\ V_\pi$ can be considered
as faithful representations of the crossed product. Thus there is an isomorphism
between $C^*(\2F, V)$ and $C^*(\pi(\2F), V_\pi)$ which maps $F\in\2F$ into $\pi(F)$
and $V$ into $V_\pi$. \qed
\bdefin DHR-Representations and transportable endomorphisms are called even iff they 
satisfy a) and b) of Lemma \ref{c-endos}, respectively. \edefin
We have thus singled out a class of representations which gives rise to localized
endomorphisms of the field algebra $\2F$. But this class is still too large in the
sense that unitarily equivalent even representations need not be inner equivalent.
Let $\rho$ be an even endomorphism of $\2F$, localized in $\2O$. Then
$\sigma=Ad_{UV}\circ\rho$ with $U\in\2F_-(\2O)$ is even and equivalent to $\rho$ as a
representation, but $(\rho,\sigma)\cap\2F=\{0\}$, which precludes an extension of 
the DHR analysis of permutation statistics etc. Furthermore, $\rho$ and $\sigma$, 
although they are equivalent as representations of $\2F$, restrict to inequivalent
endomorphisms of $\2F_+$. This observation leads us to confine our attention to the
following class of representations.
\bdefin An even DHR representation of $\2F$ is called bosonic if it restricts to a
bosonic DHR representation (in the conventional sense) of the even subnet $\2F_+$. 
\edefin
A better understanding of this class of representations is gained by the following
lemma.
\blemma There is a one-to-one correspondence between the equivalence classes of
boson\-ic even DHR representations of $\2F$ and bosonic DHR representations of $\2F_+$;
that is, equivalent bosonic even DHR representations of $\2F$ restrict to equivalent 
bosonic DHR representations of $\2F_+$. Conversely, every bosonic DHR representation of 
$\2F_+$ extends uniquely to a bosonic even DHR representation of $\2F$. \elemma
\rem It will become clear in Thm.\ \ref{214} that nothing is lost by considering only
representations which restrict to bosonic sectors of $\2F_+$. \\
\prf Clearly, the restriction of a bosonic even DHR representation of $\2F$ to $\2F_+$
is a bosonic DHR representation. Let $\rho, \sigma$ be irreducible even DHR morphisms of
$\2F$, localized in $\2O$, and let $T\in(\rho,\sigma)$. Twisted duality implies
$T\in\2F(\2O)^t$, i.e.\ $T=T_++T_-V$ where $T_\pm\in\2F_\pm$. Now both sides of
\be \sigma(F)= T_+\rho(F)T_+^*+T_-\alpha_k\circ\rho(F)T_-^*
   + T_+\rho(F)VT_-^* + T_-V\rho(F)T_+^* \ee
must commute with $\alpha_k$. The first two terms on the right hand side obviously
having this property, we obtain $T_+\rho(F)VT_-^* + T_-V\rho(F)T_+^*=0\ \forall F\in\2F$.
For $F=F^*$ this reduces to $T_+\rho(F)T_-^*=0$, which can be true only if $T_+=0$ or
$T_-=0$ since $\rho$ is irreducible. The case $T=T_-V$ is ruled out by the requirement
that the restrictions of $\rho$ and $\sigma$ to $\2F_+$ are both bosonic. Thus we
conclude that $T\in\2F_+(\2O)$ and the restrictions $\rho_+$ and $\sigma_+$ are
equivalent.

As to the converse, a bosonic DHR representation $\pi_+$
of $\2F_+$ gives rise to a local 1-cocycle \cite{r3,r4}
in $\2F_+$, i.e.\ a mapping $z: \Sigma_1\to\2U(\2F_+)$ satisfying the cocycle 
identity $z(\del_0 c)z(\del_2 c)=z(\del_1 c),\ c\in\Sigma_2$ and the locality condition
$z(b)\in\2F_+(|b|),\ b\in\Sigma_1$. This cocycle can be used as in \cite{r4,r-aqft} to 
extend $\pi_+$ to a representation $\pi$ of $\2F$ which has all the desired properties.
We omit the details. By this construction, the extensions of equivalent representations
are equivalent, an intertwiner $T\in(\rho,\sigma)$ lifting to $\pi(T)$ on $\2H$. \qed

\btheor Let $\2F$ be a complete normal field net associated to a rational net of
observables.
Then $\2F$ does not possess non-trivial bosonic even DHR representations with finite
statistics. Equivalently, there are no non-trivial bosonic DHR representations of the 
even subalgebra $\2F_+$ with finite statistics. \label{nosect1}\etheor
\prf Assume that $\2F$ has non-trivial bosonic even DHR representations; by the lemma 
this is equivalent to the existence of bosonic sectors of $\2F_+$. For the latter the
conventional DHR analysis goes through and gives rise to a semigroup $\Sigma$ of 
endomorphisms of $\2F_+$ with permutation symmetry etc. These morphisms lift to $\2F$ 
and we can apply the DR construction to $(\2F,\Sigma)$. Since all elements of $\Sigma$ 
are bosonic, no bosonization in the sense of \cite[(3.19)]{dr2} is necessary. 
All this works irrespective of the fact that $\2F$ is not a local net since the 
fermionic fields are mere spectators. That the resulting field net again satisfies 
normal commutation relations is more or less evident since the `new' fields are purely
bosonic. Furthermore, Lemma \ref{forts1} is still true when the 
`observable net' is $\7Z_2$-graded. Now the rest of the argument works just as in
Thm.\ \ref{nosect}. \qed\\
\rems 1. In the fermionic case, the even subnet $\2F_+$ has exactly one fermionic sector.
This sector is simple and its square is equivalent to the identity, as follows from the
fact that bosonic sectors of $\2F_+$ do not exist.\\
2. At this point one might be suspicious that there exist relevant DHR-like
representations of $\2F$ which are not covered by this theorem. In particular the
restriction to bosonic 
even DHR representations was made for reasons which may appear to be purely technical and
physically weakly motivated. The next theorem shows that this is not the case.

\btheor Let $\2A$ be a rational net of observables, let $\Gamma\subset\Delta$ be a
subsemigroup of DHR morphisms containing not only
bosonic sectors and let $\2F_\Gamma$ be the incomplete $\7Z_2$-graded field net
associated with $(\2A,\Gamma)$. Then an application of the DR construction with respect
to the bosonic even morphisms $\Sigma$ of $\2F_\Gamma$, as described above, leads to a 
field net $\2F_{\Gamma,\Sigma}$ which is equivalent to the complete normal field 
net $\2F_\Delta$. \label{214}\etheor
\prf Since $\2F$ is assumed to contain fermions, every $\2F(\2O)$ contains unitaries
which are odd under $\alpha_k$, giving rise to fermionic automorphisms of $\2A$.
By composition with one of these, every irreducible endomorphism of $\2A$ can be
made bosonic. It is thus clear that it suffices to extend $\2F$ by Bose fields which
implement these bosonic sectors (more precisely, their extensions to $\2F$). The
rest of the argument goes as in the preceding subsection. \qed

It is thus the existence of bosonic sectors of the even subnet which indicates that a
fermionic field net is not complete, and only such sectors need to be considered when
enlarging the field net in order to obtain the complete field net.

\sectreset{Degenerate Sectors in $1+1$ Dimensions}
\subsection{General results on degenerate sectors}
We begin with a few easy but crucial results on the set of degenerate DHR sectors.
Let $\2O\mapsto\2A(\2O)$ be a net of observables satisfying Haag duality on the line or
in $1+1$ dimensional Minkowski space. (For remarks on the duality assumption cf.\ the 
end of the next subsection.) As shown in \cite{frs1}, with each pair of localized 
endomorphisms there are associated two a priori different statistics operators 
$\ve(\rho,\eta),\ve(\eta,\rho)^*\in(\rho\eta,\eta\rho)$. 
\bdefin[\cite{khr2}] Two DHR sectors have trivial monodromy iff the corresponding
morphisms satisfy $\ve(\rho,\eta)$$=\ve(\eta,\rho)^*$ or, equivalently,
$\ve_M(\rho,\eta)=\11$ (this is independent of the
choice of $\rho,\eta$ within their equivalence classes).
A DHR sector is degenerate iff it has trivial monodromy with all sectors
(it suffices to consider the irreducible ones). \label{d1}\edefin
A convenient criterion for triviality of the monodromy of two morphisms is given by the 
following
\blemma Let $\rho,\eta$ be irreducible localized endomorphisms. Furthermore, let 
$\eta_L,\eta_R$ be equivalent to $\eta$ localized to the spacelike left and right of 
$\rho$, respectively, with the (unique up to a constant) intertwiner
$T\in(\eta_L,\eta_R)$. Then triviality of the monodromy $\ve_M(\rho,\eta)$ is
equivalent to $\rho(T)=T$. \label{l2}\elemma
\prf Using the intertwiners $T_{L/R}\in(\eta,\eta_{L/R})$, the statistics operators
are given by $\ve(\rho,\eta)=T_L^*\rho(T_L)$ and $\ve(\eta,\rho)=T_R^*\rho(T_R)$.
The monodromy operator is given by $\ve_M(\rho,\eta)=T_L^*\rho(T_L T_R^*)T_R$.
Thus, $\ve_M(\rho,\eta)=\11$ is equivalent to $\rho(T_L T_R^*)=T_L T_R^*$.
The proof is completed by the observation that $T_L T_R^*$ equals $T^*$ up to a phase. 
\qed\\
At first sight one might be tempted to erroneously conclude from this lemma that there 
is no nontrivial braid statistics as follows: The above charge transporting 
intertwiner $T$ commutes with $\2A(\2O)$ which, appealing to Haag duality, implies that 
it is contained in the algebra of the spacelike complement $\2O'$. On this algebra every
morphism localized in $\2O$ acts trivially, so that the lemma implies permutation group 
statistics. The mistake in this argument is, of course, that $T$ is contained in the
weakly closed algebra $\2R(\2O')\equiv\2A(\2O')''=\2A(\2O)'$ but not necessarily in the 
$C^*$-subalgebra $\2A(\2O')$ of the quasilocal algebra $\2A$. It is only the latter
on which $\rho$ is known to act trivially. 

\blemma A reducible DHR representation $\pi$ is degenerate iff all irreducible
subrepresentations are degenerate. \label{reduc}\elemma
\prf Let $\rho$ be equivalent to $\pi$ and localized in $\2O$, decomposing into 
irreducibles according to $\rho=\sum_{i\in I}V_i\, \rho_i(\cdot)\,V_i^*$. That is, the 
$\rho_i$ are localized in $\2O$ and $V_i\in\2A(\2O)$ with 
$V_i^*V_j=\delta_{i,j}\11$ and $\sum_iV_iV_i^*=\11$.
By Lemma \ref{l2}, $\pi$ is degenerate iff $\rho(T)=T$ for every unitary
intertwiner between  (irreducible) morphisms $\sigma,\sigma'$ which are 
localized in the two different connected components of $\2O'$. Now, 
$\rho(T)=\sum_i V_i\,\rho_i(T)\,V_i^*$ equals $T$ iff `all matrix elements are equal',
i.e.\ $V_j^*\,T\,V_k=\delta_{j,k}\,\rho_j(T)\ \forall j,k\in I$. But since
$T\in\2A(\2O)'$ the left hand side equals $TV_j^*V_k=T\delta_{j,k}$ which leads to 
the necessary and sufficient condition $\rho_j(T)=T\ \forall j\in I$, which in turn
is equivalent to all $\rho_i$ being degenerate. \qed

\blemma Let $\Delta_D$ be the set of all degenerate morphisms with finite statistical
dimension. Then $(\Delta_D,\ve)$ is a permutation symmetric, specially directed semigroup
with subobjects and direct sums. \label{trivm}\elemma
\prf Let $\rho_1,\rho_2$ be degenerate, i.e.\ $\ve_M(\rho_i,\sigma)=\11\ \forall\sigma$.
Due to the identities \cite{frs2} 
\bea \ve(\rho_1\rho_2,\sigma) &=& \ve(\rho_1,\sigma)\rho_1(\ve(\rho_2,\sigma)), 
  \label{braid1}\\
 \ve(\sigma,\rho_1\rho_2) &=& \rho_1(\ve(\sigma,\rho_2))\ve(\sigma,\rho_1) 
\label{braid2}\eea
we have
\be \ve_M(\rho_1\rho_2,\sigma)=\ve(\rho_1\rho_2,\sigma)\ve(\sigma,\rho_1\rho_2)=
  \ve(\rho_1,\sigma)\rho_1(\ve(\rho_2,\sigma)\ve(\sigma,\rho_2))\ve(\sigma,\rho_1)
  =\11. \ee
Thus $\Delta_D$ is closed under composition. By the preceding lemma the direct sum
of degenerate morphisms is degenerate, and every irreducible morphism contained in a
degenerate one is again degenerate. That $(\Delta_D,\ve)$ is specially directed in the
sense of \cite[Sect.\ 5]{dr1} follows as in \cite[Lemma 3.7]{dr2} from the fact that the
degenerate sectors have permutation group statistics. \qed

\subsection{Proof of a Conjecture by Rehren}\label{rehren}
In Sect.\ 2 we proved that the DR field net corresponding to a rational theory $\2A$ in 
$\ge 2+1$ dimensions does not have DHR sectors (with finite statistics). The
dimensionality of spacetime entered in the arguments only insofar as it implies
permutation group statistics. By the results of the preceding subsection it is now clear
that we can proceed as in Sect.\ 2, restricting ourselves to the degenerate sectors. 
More concretely, 
we apply the spatial version \cite[Cor.\ 6.2]{dr1} of the construction of the crossed 
product to the quasilocal observable algebra and the semigroup $\Delta_D$ of degenerate 
sectors with finite statistics. As the proofs in \cite[Sect.\ 3]{dr2} were given for 
$\ge 2+1$ spacetime dimensions it seems advisable to reconsider them in order to be on
the safe side, in particular as far as (twisted) duality for the field net is concerned.
\bprop Let $\2F$ be the spatial crossed product \cite[Cor.\ 6.2]{dr1} of $\2A$ by 
$(\Delta,\ve)$ where $\Delta$ is as in Lemma \ref{trivm}, and let $\2F(\2O)$ be defined 
as in the proof of \cite[Thm. 3.5]{dr2}. Then $\2O\mapsto\2F(\2O)$ is a normal field 
system with gauge symmetry and satisfies twisted duality. \label{p9}\eprop
\prf The proof of existence in \cite[Thm. 3.5]{dr2} holds unchanged as it relies 
only on algebraic arguments independent of the spacetime dimension. The same holds for 
\cite[Thm. 3.6]{dr2} with the possible exception of the argument leading to twisted 
duality on p. 73. The latter boils down algebraically to the identity 
$\2F(\2O)'\cap G'=\pi(\2A(\2O'))^-,\ \2O\in\2K$. Finally, the proof of this formula in 
\cite[Lemma 3.8]{dr2} is easily verified to be correct in $1+1$ dimensions, too,
provided $\2A\restr\2H_0$ satisfies duality. In our case this is true by assumption. 
\qed\\
\rems 1. The reader who feels uneasy with these few remarks is encouraged to study the 
proofs of \cite[Thms. 3.5, 3.6]{dr2} himself, for it would make no sense to reproduce 
them here.\\
2. It may be confusing that in theories with group symmetry satisfying the split 
property for wedges (SPW), Haag duality for a field net $\2F$ and the $G$-fixpoint net
$\2A$ (in the vacuum sector) are in fact incompatible \cite{mue1}. The SPW has been 
verified for massive free scalar and Dirac fields and is probably true in all reasonable
{\it massive} theories. On the other hand, a net of observables which satisfies Haag 
duality and the SPW does not admit DHR sectors anyway \cite{mue3}. In view of this
result,  we implicitly assume in this section that the observables do {\it not} satisfy 
the SPW. The point is that one must be careful to distinguish between conformally
covariant or at least massless theories, with which we are concerned here, and massive
theories since the scenarios are quite different.

\btheor Let $\Delta_D$ be the set of all degenerate morphisms with finite statistics,
corresponding to only finitely many sectors. If $\Delta_D$ is purely bosonic, the local
field net $\2F$ constructed from $\2A$ and $\Delta_D$ does not have 
degenerate sectors with finite statistics. If $\Delta_D$ contains fermionic sectors, the
normal field net $\2F$ does not have degenerate bosonic even sectors with finite 
statistics. Equivalently, the even subnet has no degenerate bosonic sectors with finite
statistics. \label{no-deg2}\etheor
\prf The proofs of Thms.\ \ref{nosect}, \ref{nosect1} are valid also in the 
2-dimensional situation since neither the argument of Lemma \ref{forts1} on the 
extendibility of local symmetries nor the uniqueness result of \cite[Thm. 3.5]{dr2}
require any modification. \qed

This result, which was conjectured by Rehren in \cite{khr2}, is quite interesting and
potentially useful for the analysis of superselection structure in $1+1$ dimensions.
It seems worthwhile to restate it in the following form.
\bcoro Every degenerate \qft\ in $1+1$ dimensions (in the sense that there are 
degenerate sectors, which in the rational case is equivalent to non-invertibili\-ty of 
$S$) arises as the fixpoint theory of a non-degenerate theory under the action of a 
compact group of inner symmetries. That is, all degenerate theories are orbifold 
theories in the sense of \cite{dvvv}. \ecoro

Now we indicate how the preceding arguments have to be changed in the case of conformal
theories. We first remark that everything we have said about theories in $1+1$ dimension
remains true for theories on the line. In the rest of this subsection we consider
chiral theories on the circle \cite{frs2}, which are defined by associating a von 
Neumann algebra $\2A(I)$ to each interval $I$ which is not dense in $S^1$. The 
`spacetime' symmetry is given by the M\"obius group $PSU(1,1)\cong PSL(2,\7R)$.
The set of intervals not being directed, the quasilocal algebra must be replaced 
by the universal algebra of \cite{frs2} which has a non-trivial center due to the 
non-simply connectedness of $S^1$. Triviality of the center of the $C^*$-algebra $\2A$ 
was, however, an essential requirement for the Doplicher-Roberts analysis, in particular
\cite{dr4}. One may try to eliminate this condition, but we prefer another approach.
We begin by restricting the theory to the punctured circle, i.e.\ the line, for which 
one has the conventional quasilocal algebra $\2A$ which is simple. The following result 
shows that the restriction of generality which this step seems to imply -- since 
Haag duality on $\7R$ is equivalent to strong additivity, which does not hold for all 
theories -- is only apparent. 
\bprop Given a chiral conformal precosheaf the localized endomorphisms of the algebra
$\2A$ (corresponding to the punctured circle) and their statistics can be defined without
assuming duality on the real line. In the case without fermionic degenerate sectors the 
DR construction can be applied to the (M\"obius covariant) degenerate sectors with 
finite statistics and yields a conformal precosheaf $\2F$ on $S^1$ which is M\"obius 
covariant with positive energy and does not have degenerate sectors (with finite 
statistics). \eprop
\prf Let $\2A$ be the quasilocal algebra obtained after removing a point at infinity.
By the results of \cite[Sect.\ 5]{frs2} the semigroup $\Delta\subset\mbox{End}\,\2A$ of 
localized endomorphisms with braiding can be defined without assuming duality on the
line. We can thus apply the DR construction to $\2A$ and $\Delta_D$ and obtain a 
field net $\7R\supset I\mapsto\2F(I)$, but we clearly cannot hope to prove Haag duality
on $\7R$. Before we can prove duality on $S^1$ (as it holds for the observables)
we must define the local algebras for intervals which contain the point at infinity. 
Due the conformal spin-statistics theorem \cite{gl2} a covariant bosonic degenerate 
sector, having statistics phase 1, is in fact covariant under the uncovered M\"obius 
group $PSL(2,\7R)$, and consequentially also the extended theory $\2F$ is M\"obius 
covariant with positive energy. This fact can be used to define the missing local 
algebras and to obtain a conformal precosheaf on $S^1$. Then the abstract results of 
\cite{bgl,fro3} apply and yield Haag duality on $S^1$. The argument to the effect 
that the extended theory is non-degenerate works as above. \qed

\subsection{Relating the Superselection Structures of $\2A$ and $\2F$}
In the preceding subsection we have seen that whenever there are degenerate sectors
one can construct an extended theory which is non-degenerate. The larger theory has a
group symmetry such that the original theory is reobtained by retaining only the
invariant operators. Equivalently, all degenerate theories are orbifold theories. By 
this result, a general analysis of the superselection structure in $1+1$ dimensions
may begin by considering the non-degenerate case. It remains, however, to clarify the
relation between the superselection structures of the degenerate theory and the extended
theory. This will not be attempted here, but we will provide some results going in this
direction.

\blemma All irreducible morphisms contained in the product of a degenerate morphism and
a non-degenerate morphism are non-degenerate. \elemma
\prf The fact that the composition of degenerate morphisms yields a sum of degenerate 
morphisms can be expressed in terms of the fusion coefficients $N_{ij}^k$ as
\be \mbox{$i$ and $j$ degenerate, $k$ non-degenerate} \ \impl\ N_{ij}^k=0.\ee
By Frobenius reciprocity $N_{ij}^k=N_{i\bar{k}}^{\bar{\jmath}}$ this implies 
\be \mbox{$i$ and $j$ degenerate, $k$ non-degenerate} \ \impl\ N_{ik}^j=0 .\ee
(We have used the fact that the conjugate $\bar{\rho}$ is degenerate iff $\rho$ is
degenerate.) \qed\\
\rem This simple fact may be interpreted by saying that the set of non-degenerate sectors
is acted upon by the set of degenerate ones, i.e.\ a group dual $\hat{G}$. If $G$ is
abelian, $K=\hat{G}$ is itself an abelian group and we are in the situation studied, 
e.g., in \cite{fss}. The result of the preceding subsection thus constitutes a
mathematically rigorous though rather abstract solution of the field identification 
problem \cite{sy}, cf.\ also the remarks in the concluding section.

Being able to apply the DR construction also in $1+1$ dimensions provided we consider
only semigroups of degenerate endomorphisms, we are led to reconsider Lemma
\ref{forts2} concerning the extension of
localized endomorphisms of the observable algebra to the field net. The construction 
given in Sect.\ 2 can not be used for the extension of non-degenerate morphisms $\eta$
since we do not have a complete field net at our disposal. There are at least two
approaches to the problem which do not rely on the existence of a complete field net.
The first one \cite[Prop. 3.9]{lr} uses the inclusion theory of von Neumann factors
which, however, we want to avoid in this work since a proof of factoriality of the local
algebras exists only for conformally covariant QFTs but not for general theories.
Another prescription was given by Rehren \cite{khr1}. The claim of uniqueness made there
has, however, to made more precise. Furthermore, it is not completely trivial to 
establish the existence part. Fortunately, both of these questions can be clarified in a
relatively straightforward manner by generalizing results by Doplicher and Roberts.
In \cite[Sect.\ 8]{dr1} they considered a similar extension problem, namely the
extension of automorphisms of $\2A$ to automorphisms of $\2F$ commuting with the
gauge group. The application that these authors had in mind was the extension of 
spacetime symmetries to the field net \cite[Sect.\ 6]{dr2} under the provision that the 
endomorphisms implemented by the fields are covariant. For a morphism $\rho\in\Delta$ 
the inner endomorphism of $\2F$ which extends $\rho$ will also be denoted by $\rho$. 
\blemma Let $\2B$ be the crossed product \cite{dr1} of the $C^*$-algebra $\2A$ with
center $\7C\11$ by the permutation symmetric, specially directed semigroup $(\Delta,\ve)$
of endomorphisms and let $G$ be the corresponding gauge group. Let $\Gamma$ be a 
semigroup of unital endomorphisms of $\2A$. Then 
there is a one-to-one correspondence between actions of $\Gamma$ on $\2B$ by unital
endomorphisms $\tilde{\eta}$ which extend $\eta\in\Gamma$ and commute elementwise with 
$G$ and mappings $(\rho,\eta)\mapsto W_\rho(\eta)$ from $\Delta\times\Gamma$ to 
unitaries of $\2A$ satisfying
\bea W_\rho(\eta) & \in & (\rho\eta,\eta\rho),  \label{w1}\\
   W_{\rho'}(\eta)TW_\rho(\eta)^* &=& \eta(T), \ \ T\in(\rho,\rho'),  \label{w2} \\
   W_{\rho\rho'}(\eta) &=& W_\rho(\eta)\, \rho(W_{\rho'}(\eta)),  \label{w3}\\
   W_\rho(\eta\eta') &=& \eta(W_\rho(\eta'))W_\rho(\eta)  \label{w4} \eea
for all $\eta,\eta'\in\Gamma,\,\rho,\rho'\in\Delta$. The correspondence is determined by 
\be \tilde{\eta}(\psi)=W_\rho(\eta)\psi, \ \ \psi\in H_\rho,\ \rho\in\Delta, 
   \eta\in\Gamma.  \label{w5}\ee
Furthermore, if a unitary $S\in(\eta,\eta'),\ \ \eta,\eta'\in\Gamma$ satisfies
\be SW_\rho(\eta)=W_\rho(\eta')\rho(S) \ \ \forall \rho\in\Delta ,\label{stilde}\ee
then $S\in(\tilde{\eta},\tilde{\eta}')$. \label{p6}\elemma
\prf An inspection of the proofs of \cite[Thm.\ 8.2, Cor.\ 8.3]{dr1}, where groups of
automorphisms were considered, makes plain that they are valid also for the case of 
semigroups of true endomorphisms and we refrain from repeating them.
Besides $\eta$ not necessarily being onto, the only change occurred in (\ref{w1}) which 
replaces the property $W_\rho\in(\rho,\beta\rho\beta^{-1})$ which does not make sense 
for a proper endomorphism $\beta$. Given $\tilde{\eta}$ and setting
\be W_\rho=\sum_{i=1}^d \tilde{\eta}(\psi_i)\psi_i^* ,\ee
where $\psi_i,\, i=1,\ldots,d$ is a basis of $H_\rho$, it is clear that $W_\rho$
satisfies (\ref{w1}). The other properties of the $W$'s are proved as in \cite{dr1}.
As to the converse direction, we are done provided we can show that \cite[Thm. 8.1]{dr1}
concerning the extension of $\eta$ to the cross product of $\2A$ by a single
endomorphism $\rho$ generalizes to the case of $\eta$ an endomorphism. We give only that
part of the argument which differs from the one in \cite{dr1}.

Therefore let $\2A$ and $\rho$ satisfy the assumptions of \cite[Thm. 8.1]{dr1}, let
$\eta$ be an injective unital endomorphism of $\2A$ and let $W\in(\rho\eta,\eta\rho)$
satisfy \cite[(8.1), (8.2)]{dr1}.
As in \cite[Thm. 8.1]{dr1} we consider the monomorphisms of $\2A$ and $\2O_d$
into $\2A\otimes_\mu\2O_d$, defined by $\pi': A\mapsto\eta(A)\otimes_\mu\11$ and
$\zeta': \psi\mapsto W\otimes_\mu\psi,\ \psi\in H$, respectively. The calculation leading
to $\zeta'(\psi)\pi'(A)=\pi'\circ\rho(A)\zeta'(\psi)$ is correct also for $\eta$ a true
endomorphism. Furthermore, with $\zeta_1'=\zeta'\restr\2O_{SU(d)}$ we have 
$\zeta_1'(\2O_{SU(d)})\in\eta(\2A)$ thanks to the conditions on $W$ and the fact that
$\2O_{SU(d)}$ is generated by the elements $S$ and $\theta$, see \cite{dr5}. Thus 
$\eta\rho\eta^{-1}\circ\zeta_1'$ is well defined and equals $\zeta_1'\circ\sigma$, where
$\sigma$ is the canonical endomorphism of $\2O_{SU(d)}$. As in \cite{dr1} we conclude
that $\zeta'\restr\2O_{SU(d)}=\eta\circ\mu$.
Thus by the universality of $\2A\otimes_\mu\2O_d$ there is an isomorphism between 
$\2A\otimes_\mu\2O_d$ and the subalgebra generated by $\pi'(\2A)$ and $\zeta'(\psi)$.
Equivalently, there is an endomorphism $\gamma$ of $\2A\otimes_\mu\2O_d$ such that
\bea \gamma(A\otimes_\mu\11) &=& \eta(A)\otimes_\mu\11,\ \ \ A\in\2A, \\
  \gamma(\11\otimes_\mu\psi) &=& W\otimes_\mu\psi,\ \ \ \psi\in H. \eea
Now the rest of the proof goes exactly as in \cite[Thm. 8.1]{dr1}, i.e.\ after factoring
out the ideal $J_\phi$ we obtain an endomorphism $\tilde{\eta}$ of the crossed product
$\2B=(\2A\otimes_\mu\2O_d)/J_\phi$ which commutes with the action of the gauge group $G$.

Now let $S\in(\eta,\eta')$ satisfy (\ref{stilde}). Then for $\psi\in H_\rho$ we have
\be S\tilde{\eta}(\psi)=SW_\rho(\eta)\psi=W_\rho(\eta')\rho(S)\psi
   =W_\rho(\eta')\psi S=\tilde{\eta}'(\psi)S .\ee
Since $\tilde{\eta},\tilde{\eta}'$ are determined by their action on the spaces $H_\rho$
this implies $S\in(\tilde{\eta},\tilde{\eta}')$. \qed

We are now ready to consider the wanted extensions of localized endomorphisms.
Motivated by Lemma \ref{forts2} where we had (in the case of bosonic $\psi^\rho$) 
\be \hat{\eta}(\psi^\rho)=\sum_i \psi^\eta_i\,\psi^\rho\,\psi_i^{\eta*} =
  \left(\sum_{i,j} \psi^\eta_i\psi_j^\rho\psi_i^{\eta*}\psi_j^{\rho*}\right)\,\psi^\rho
   = \ve(\rho,\eta)\,\psi^\rho, \ee
we appeal to the preceding lemma with $W_\rho(\eta)=\ve(\rho,\eta)$. 

\bprop Let $\2O\mapsto\2F(\2O)$ be the field net obtained via the Doplicher-Roberts
construction from the algebra $\2A$ of observables and a semigroup of degenerate 
morphisms, closed under direct sums, subobjects and conjugates.
Then every localized (unital) endomorphism $\eta$ of $\2A$ extends to a localized 
endomorphism $\tilde{\eta}$ of $\2F$ commuting with the action of the gauge group. 
If $\eta$ is localized in $\2O$ the same holds for $\tilde{\eta}$. Every 
$S\in(\eta,\sigma)$ lifts to $S\in(\tilde{\eta},\tilde{\sigma})$. \label{forts3}\eprop
\prf We set $W_\rho(\eta)=\ve(\rho,\eta)$ and verify the requirements 
(\ref{w1}-\ref{w4}). Obviously (\ref{w1}) is fulfilled by definition of
the statistics operator. (\ref{w3}) and (\ref{w4}) follow from (\ref{braid1}) and
(\ref{braid2}), respectively.
Finally, (\ref{w2}) is just $\ve(\rho',\eta)T=\eta(T)\ve(\rho,\eta)$ which holds for
$T\in(\rho,\rho')$. The statement on the localizations follows from the fact that 
$\ve(\rho,\eta)=\11$ if $\rho,\eta$ are spacelike localized, since $\rho$ is degenerate.
Finally, with
$S\in(\eta,\sigma)$ one has $\ve(\rho,\eta)S=\rho(S)\ve(\rho,\sigma)$ such that the 
condition (\ref{stilde}) is satisfied. Thus $S\in(\tilde{\eta},\tilde{\sigma})$. \qed\\
\rems 1. The above result is unaffected if the field net is fermionic. In this case the
identitity $\tilde{\eta}\circ\alpha_k=\alpha_k\circ\tilde{\eta}$ where $k\in G$ is
the grading element (which distinguishes bosonic and fermionic fields) shows that
$\tilde{\eta}$ leaves the statistics of fields invariant. In fact, this observation 
provided the motivation for introducing the notion of even DHR sectors in Sect.\
\ref{22}.\\
2. As already remarked, an alternative proof can be given using \cite[Prop.\ 3.9]{lr}. 
In this approach the localization of the extended morphism in a double cone follows 
since the morphism $\rho$ appearing there is the restriction to $\2A(\2O)$ of the 
canonical endomorphism $\gamma$ from $\2F(\2O)$ into $\2A(\2O)$ for some $\2O\in\2K$. 
But in the situation at hand we have 
$\rho\cong\bigoplus_{\xi\in\Delta_{\mbox{deg}}} d_i\rho_\xi$, thus $\rho$ is 
degenerate. \\
3. In principle,
the construction of the field algebra works for every family of sectors with permutation
group statistics which is closed under direct sums and subobjects. 
As emphasized by Rehren \cite{khr1}, the extension $\tilde{\eta}$ is localized in
a double cone only if the charged fields in $\2F$ correspond to {\it degenerate} sectors,
for otherwise $\ve(\rho,\eta)=\11$ holds only if $\rho$ is localized to the right of
$\eta$ (or left, if $\ve(\eta,\rho)^*$ is used).

In the special case where $\eta$ is an automorphism, the extension $\tilde{\eta}$ can
be defined via \cite[Thm. 8.2]{dr1}, using $W_\rho(\eta)$ as above. Clearly, 
$\tilde{\eta}$ is irreducible since it is an automorphism. In general, however, the 
extension $\tilde{\eta}$ will not be irreducible. Rehren's description 
\cite{khr1} of the relative commutant can also be proved rigorously by adapting 
earlier results \cite[Lemma 5.1]{dr3}.
\blemma The relative commutant $\2F\cap\tilde{\eta}(\2F)'$ is generated as a closed
linear space by sets of the form $(\rho\eta,\eta)H_\rho,\ \rho\in\Delta$.
\label{relc}\elemma
\prf By twisted duality, an operator in $\tilde{\eta}(\2F)'$ is contained in
$\2F(\2O)^t$, where $\2O$ is the localization region of $\eta$. Due to 
$\2F\cap\2F^t=\2F_+$, the self\/intertwiners of $\tilde{\eta}$ in $\2F$ are bosonic.
Obviously, $(\rho\eta,\eta)\psi,\ \psi\in H_\rho$ is in $\2F\cap\eta(\2A)'$. Now, just 
as $\eta$, so can $\rho\eta$ be extended to an endomorphism $\widetilde{\rho\eta}$
of $\2F$ by the proposition. Furthermore, $T\in(\rho\eta,\eta)$ lifts to an intertwiner
between $\tilde{\rho\eta}$ and $\tilde{\eta}$. Thus $(\rho\eta,\eta)\psi^\rho$ is in 
$\2F\cap\tilde{\eta}(\2F)'$. As to the converse, $\tilde{\eta}(\2F)'\cap\2F$ is globally 
stable under the action of $G$ since $\tilde{\eta}$ commutes with $G$. Thus 
$\tilde{\eta}(\2F)'\cap\2F$ is generated linearly by its irreducible tensors under $G$.
If $T_1, \ldots, T_d$ is such a tensor from $\2F\cap\tilde{\eta}(\2F)'$, then
there is a multiplet $\psi_i, i=1, \ldots, d$ of isometries in $\2F$
and transforming in the same way, since the field algebra has full Hilbert $G$-spectrum.
With $X=\sum_{i=1}^d T_i\psi_i^*\in\2F^G$ we have $T_i=X\psi_i$ and we must prove
$X\in(\rho\eta,\eta)$. Now $T_iF=FT_i$ for $F\in\tilde{\eta}(\2F)$ implies
\be X\psi_i \, \tilde{\eta}(F) = X\rho(\tilde{\eta}(F))\psi_i=\tilde{\eta}(F)X\psi_i, 
  \ \ F\in\2F, i=1,\ldots,d .\ee
Multiplying the second identity with $\psi_i^*$ and summing over $i$ we obtain 
$X\rho(\tilde{\eta}(F))=\tilde{\eta}(F)X, \ \ F\in\2F$ since 
$\sum_i\psi_i\psi_i^*=\11$ by construction of the field algebra. Thus 
$X\in(\rho\eta,\eta)$. \qed
\bcoro $\tilde{\eta}$ is irreducible iff the endomorphism $\eta\bar{\eta}$ of $\2A$
does not contain a nontrivial morphism $\rho\in\Delta$. \ecoro
\prf By the lemma, the existence of a morphism $\rho\in\Delta$ with 
$(\eta\rho,\eta)\ne\{0\}$ is necessary and sufficient for the nontriviality of 
$\2F\cap\tilde{\eta}(\2F)'$. But by Frobenius reciprocity, $\eta\rho\succ\eta$ is 
equivalent to $\eta\bar{\eta}\succ\bar{\rho}$. \qed\\
\rem The irreducible endomorphisms obtained by decomposing an extension 
$\tilde{\eta}$ are even, provided we use bosonic isometric intertwiners. This can
always be done as the relative commutant is contained in $\2F_+$ by Lemma \ref{relc}.

In the above considerations we had to assume, for the technical reasons explained in
Sect.\ 2, that there are only finitely many degenerate sectors. There was, however, no
restriction on the number of non-degenerate sectors. We conclude with an important
observation concerning rational theories.
\bprop Let $\2A$ have finitely many DHR sectors (degenerate and non-degene\-rate) with
finite statistics. Then the extended theory $\2F$ has only finitely many sectors with
finite statistics (all of which are non-degenerate by Thm.\ \ref{no-deg2}). 
\label{rat}\eprop
\prf As a consequence of Prop.\ \ref{forts3} we have 
$\widetilde{\eta_1\oplus\eta_2}\cong\tilde{\eta}_1\oplus\tilde{\eta}_2$, since
the intertwiners $V_i\in(\eta_i,\eta_1\oplus\eta_2), i=1,2$ are in 
$(\tilde{\eta}_i,\tilde{\eta}_1\oplus\tilde{\eta}_2)$.
Thus every irreducible $\2F$-sector $\beta$ which is contained in the extension 
$\tilde{\eta}$ of an $\2A$-sector $\eta$ is already contained in the extension of an 
irreducible $\eta'$. As observed in \cite{khr1} the statistical dimensions of 
$\eta, \tilde{\eta}$ satisfy $d_\eta=d_{\tilde{\eta}}$. (This follows from the existence
of a left inverse $\tilde{\phi}_{\tilde{\eta}}$ which extends $\phi_\eta$ and the fact 
that $\ve(\eta,\eta)=\ve(\tilde{\eta},\tilde{\eta})$ again by Prop.\ \ref{forts3}.) Thus
$\tilde{\eta}$ decomposes into finitely many irreducible $\2F$-sectors. Since by
assumption the number of irreducible $\2A$-sectors is finite we obtain only finite many 
irreducible $\2F$-sectors by inducing from $\2A$. The claim follows provided we can show
that every irreducible $\beta\in\Delta_\2F$ is contained in some $\tilde{\eta}$. This is
true by the observation after \cite[Thm.\ 3.21]{be}, but we prefer to give a direct 
argument. 

For $\beta\in\Delta_\2F(\2O)$ also 
$\beta_g=\alpha_g\circ\beta\circ\alpha_g^{-1}, g\in G$ is localized in $\2O$ and
transportability is easy. Let now $\{V_g, g\in G\}$ be a multiplet of 
isometries in $\2F(\2O)$ satisfying $V_g^*V_h=\delta_{g,h}\11$, $\sum_gV_gV_g^*=\11$
and $\alpha_h(V_g)=V_{hg}$. Such a multiplet exists since the action of $G$ on
$\2F(\2O)$ by construction has full Hilbert spectrum. Then
\be \ol{\beta}(\cdot):=\sum_{g\in G} V_g\,\beta_g(\cdot)\,V_g^* \ee
commutes with the action of $G$ and thus restricts to a transportable 
endomorphism of $\2A$ which is localized in $\2O$. Now $\ol{\beta}$ can be considered 
as an extension to $\2F$, commuting with $G$, of $\ol{\beta}\restr\2A$ and by
Lemma \ref{p6} there exists a family $\{W_\rho(\ol{\beta}\restr\2A), \rho\in\Delta\}$ 
satisfying (\ref{w1})-(\ref{w3}) and the boundary condition 
$W_\rho(\ol{\beta}\restr\2A)=\11$ for $\rho\in\Delta_\2A(\2O')$. Since $\rho$ is 
degenerate the unique solution to these equations is 
$W_\rho(\ol{\beta}\restr\2A)=\ve(\rho,\ol{\beta}\restr\2A)$. This implies 
$\widetilde{\ol{\beta}\restr\2A}=\ol{\beta}$ and in view of 
$\beta=\beta_e\prec\ol{\beta},\ \ \beta$ is contained in
the extension to $\2F$ of the $\2A$-sector $\ol{\beta}\restr\2A$. \qed\\
\rem The connection with the argument in \cite[Thm.\ 3.21]{be} is provided
by the observation that the expression $\gamma=\sum_g V_g\,\alpha_g(\cdot)\,V_g^*$ built 
with the above $V's$ is nothing but the extension \cite[Cor.\ 3.3]{lr} to $\2F$ of a 
canonical endomorphism from $\2F(\2O)$ into $\2A(\2O)$. Furthermore, 
$\ol{\beta}\restr\2A$ coincides with the canonical restriction 
$\sigma_\beta:=\gamma\circ\beta\restr\2A$ of \cite{lr,be}. Relying on our version of the
induction procedure (Lemma \ref{p6}), the above proof is clearly independent of the 
inclusion theory of von Neumann factors used in \cite{lr,be} and thus more in the
spirit of DHR/DR theory to which the assumption of local factoriality is
alien. (The main application of our considerations being to conformal theories where 
local factoriality is automatic, this added generality admittedly is not very important.)

\sectreset{A Sufficient Criterion for Non-Degeneracy}
The results of this section depend on an additional axiom, the split property.
\begin{defin} An inclusion $A\subset B$ of von Neumann algebras is split \cite{dl}
if there exists a type-I factor $N$ such that $A\subset N\subset B$. A net of 
algebras satisfies the `split property for double cones' 
if the inclusion $\2A(\2O_1)\subset\2A(\2O_2)$ is split whenever 
$\2O_1\subset\subset\2O_2$, i.e.\ the closure of $\2O_1$ is contained in the interior 
of $\2O_2$. \end{defin}
The importance of this property derives from the fact \cite{dal,dl} that it is
equivalent
to the following formulation: For each pair of double cones $\2O_1\subset\subset\2O_2$
the algebra $\2A(\2O_1)\vee\2A(\2O_2)'$ is unitarily equivalent to the tensor product
$\2A(\2O_1)\otimes\2A(\2O_2)'$. It is believed that this form of the split property is
satisfied in all reasonable \qfts.

In the rest of this section we will give a sufficient criterion for the {\it absence}
of degenerate sectors. The following result is independent of the number of spacetime
dimensions.
\bprop Let $\2O\mapsto\2A(\2O)$ be a net of observables fulfilling
Haag duality and the split property for double cones. Let $\rho$ be an endomorphism
of the quasilocal algebra $\2A$ which is localized in $\2O$ and acts identically on the 
relative commutant $\2A(\hat{\2O})\cap\2A(\2O)'$ whenever $\hat{\2O}\supset\supset\2O$.
Then $\rho$ is an inner endomorphism of $\2A$, i.e.\ a direct sum of copies of the 
identity morphism. \label{do-lemma}\eprop
\prf Choose double cones $\2O_1,\2O_2$ fulfilling
$\2O\subset\subset\2O_1\subset\subset\2O_2\subset\subset\hat{\2O}$. Thanks to the
split property there exist type I factors $M_1,M_2$ such that
\be \2A(\2O)\subset\2A(\2O_1)\subset M_1\subset\2A(\2O_2)\subset M_2\subset
   \2A(\hat{\2O})  .\ee
We first show $\rho(M_1)\subset M_1$. If $A\in M_1$ we have
$\rho(A)\in\2A(\2O_2)$. Due to $\2A(\2O)\subset M_1$ and the premises, $\rho$ acts 
trivially on $M_1'\cap\2A(\hat{\2O})$. Thus
\be \rho(A)\in (M_1'\cap\2A(\hat{\2O}))'\cap M_2 
   \subset(M_1'\cap M_2)'\cap M_2 = M_1 .\ee
The last identity follows from $M_1, M_2$ being type I factors.
Thus $\rho$ restricts to an endomorphism of $M_1$.
Now every endomorphism of a type I factor is inner \cite[Cor.\ 3.8]{lo1}, i.e.\ 
there is a (possibly infinite) family of isometries $V_i\in M_1,\ i\in I$ with 
$V_i^* V_j=\delta_{i,j},\sum_{i\in I}V_i V_i^*=\11$ such that 
\be \rho\restr M_1=\eta(A)\equiv\sum_{i\in I}V_i\ \cdot \ V_i^*  \label{inner}\ee
(The sums over $I$ are understood in the strong sense.)
Now by (\ref{inner}) and the premises, $\eta\restr \2A(\2O)'\cap M_1=id$, implying
$V_i\in(\2A(\2O)'\cap M_1)'\cap M_1=\2A(\2O)\ \forall i\in I$. 
Therefore $\rho=\eta$ on $\2A(\2O_1)$ and
$\rho=\eta=id$ on $\2A(\2O')$. In order to prove $\rho=\eta$ on all of $\2A$
it suffices to show $\rho(A)=\eta(A) \ \forall A\in\2A(\2O_2)$ where we may, of
course, assume $\2O_2\supset\2O_1$. For the moment we take for granted that
\be \2A(\2O_2)=\2A(\2O_1)\vee(\2A(\2O_2)\wedge\2A(\2O)') .\label{o1o2}\ee
Having just proved $\rho\restr\2A(\2O_1)=\eta$ and remarking that 
$\rho\restr\2A(\2O_2)\wedge\2A(\2O)'=id=\eta$ is true by assumption, we conclude by
local normality that $\rho\restr\2A(\2O_2)=\eta$.
In order to prove (\ref{o1o2}), apply the split property to the inclusion 
$\2O\subset\subset\2O_1$. Under the split isomorphism we have
\bea \2A(\2O) &\cong& \2A(\2O)\otimes\11, \nn\\
   \2A(\2O_i) &\cong& \2B(\2H_0)\otimes\2A(\2O_i),\ i=1,2. \label{isos}\eea
Thus $\2A(\2O)'\cong\2A(\2O)'\otimes\2B(\2H_0)$ and 
$\2A(\2O_2)\wedge\2A(\2O)'\cong\2A(\2O)'\otimes\2A(\2O_2)$, from which (\ref{o1o2})
follows at once. \qed\\
\rem The first part of the proof is essentially identical to \cite[Prop. 2.3]{dopl}. 
There it was stated only for automorphisms but the possibility of the above extension 
was remarked. In \cite{dopl} the $C^*$-version of the time-slice axiom was used to 
conclude $\rho=\eta$ on $\2A$. We have dispensed with this assumption by requiring 
triviality of $\rho$ on the relative commutant $\2A(\hat{\2O})\cap\2A(\2O)'$ for all 
$\hat{\2O}\supset\supset\2O$. For our purposes this will be sufficient.

We are now in a position to state our criterion for the absence of degenerate sectors
in $1+1$ dimensions:
\bcoro Assume in addition to the conditions of the proposition that for each pair
$\2O\subset\subset\hat{\2O}$ the algebra $\2A(\hat{\2O})\cap\2A(\2O)'$ is generated by 
the charge transporters from $\2O_L$ to $\2O_R$ (and vice versa). Here $\2O_L,\2O_R$
are the connected components of $\hat{\2O}\cap\2O'$, see the figure below. Then there 
are no degenerate sectors. More precisely, every degenerate endomorphism is inner in 
the above sense. \label{nodeg}\ecoro
\prf Due to Lemma \ref{l2}, a degenerate morphism localized in $\2O$ acts trivially on
the charge transporters between $\2O_L$ to $\2O_R$. As these are weakly dense in 
$\2A(\hat{\2O})\cap\2A(\2O)'$ by assumption and due to local normality, the morphism 
acts trivially on the relative commutant.  This is true for every 
$\hat{\2O}\supset\supset\2O$. The statement now follows from Prop.\ \ref{do-lemma}. 
\qed\\
\rems 1. In \cite{mue3} we show that a much further-reaching result can be proved if 
one requires the split property not only for double cones but also for wedge regions.
This property can, however, hold only in massive \qfts.\\
2. Admittedly the condition on the relative commutant seems difficult to verify.
One may perhaps hope that something can be said in the case of rational theories,
which have finitely many sectors.\\
3. It is likely that the condition on the relative commutant made in the corollary is
also necessary. The argument goes as follows. If there are degenerate sectors then there
is a field net $\2F$ with group symmetry such that the net $\2A$ is the restriction of 
the invariant subnet to the vacuum sector, cf.\ the next section. Assuming that the 
field net also satisfies
the split property, one can define localized implementers of the gauge group as in
\cite{bdl}. In particular, for every inclusion $\Lambda=(\2O\subset\subset\hat{\2O})$ and
every $x\in U(G)''\cap U(G)'$ one obtains an operator  
$U_\Lambda(x)\in\2A(\hat{\2O})\cap\2A(\2O)'$. We see no reason why $U_\Lambda(x)$ 
should be contained in the algebra generated by $\2A(\2O_L),\2A(\2O_R)$ and the
charge transporters.

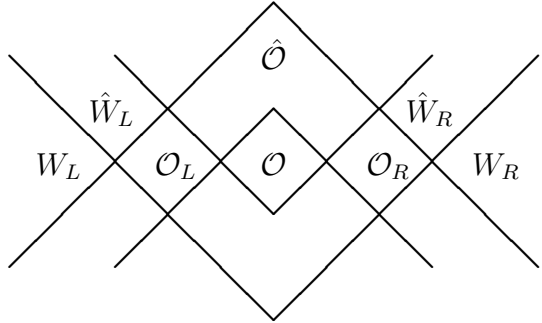
\begin{figure}
\[\ba{c}
\begin{picture}(300,150)(-150,-75)\thicklines
\put(0,60){\line(1,-1){100}}
\put(0,60){\line(-1,-1){100}}
\put(0,-60){\line(1,1){100}}
\put(0,-60){\line(-1,1){100}}

\put(0,20){\line(1,-1){60}}
\put(0,20){\line(-1,-1){60}}
\put(0,-20){\line(1,1){60}}
\put(0,-20){\line(-1,1){60}}

\put(-5,-5){$\2O$}
\put(-5,35){$\hat{\2O}$}
\put(-45,-5){$\2O_L$}
\put(35,-5){$\2O_R$}
\put(-90,-5){$W_L$}
\put(75,-5){$W_R$}
\put(-70,15){$\hat{W}_L$}
\put(50,15){$\hat{W}_R$}
\end{picture} 
\ea\]
\caption{Relative commutant of double cones}
\end{figure}

\sectreset{Summary and Outlook}
In this work we have proved two intuitively reasonable properties of the 
Doplicher-Roberts construction: The (essentially unique) complete field net which 
describes all DHR sectors has itself no localized sectors, and also it can be 
obtained from an intermediate, thus incomplete, field net by an application of the DR 
construction. Unfortunately, we have been able give a proof only under the quite
restrictive assumption that
there are only finitely many sectors, which is equivalent to finiteness of the gauge
group $G$. We emphasize that the problem consists in proving compactness of $\ol{G}$ 
in Prop.\ \ref{forts1}. If this proposition can be generalized the rest of the arguments
goes through unchanged. In any case, the complete (w.r.t.\ the DHR sectors) field net 
may still have nontrivial representations with the weaker Buchholz-Fredenhagen 
localization property. 

In a sense, the situation in low dimensions is quite similar. The degenerate sectors 
may be considered `better localized' than generic DHR sectors insofar as they arise from
{\it local} fields, in contrast to what is to be expected in the general case. 
Non-local charged fields played a role, e.g., in \cite{mue1} where, however, the 
underlying quantum symmetry was spontaneously broken. As we intend to show elsewhere, 
the symmetry breakdown encountered there is generic in massive models. As was mentioned
above, the peculiar nature of the superselection structure of massive models manifests 
itself also in an analysis which starts from the observables \cite{mue3}. For this 
reason, the considerations in Sects.\ 3 and 4 were aimed primarily at conformally 
covariant theories in $1+1$ dimensions. 

Turning to a brief discussion of open problems, the most obvious one is relaxing the
rationality assumption on the superselection structure of the observable net in the
proof of Prop.\ \ref{forts1}, which is the basis of most of our results. In trying so it
is not inconceivable that one may find counterexamples, but the author is convinced that
this cannot happen for theories with countably many sectors.

The hope expressed in Remark 2 after Cor.\ \ref{nodeg} has already been vindicated 
by reducing the relative commutant property needed there to a simple numerical identity 
which can be proved for large classes of models, cf.\ \cite{mue7}.

In Subsect.\ \ref{rehren} and Prop.\ \ref{rat} we have proven that every rational QFT in 
$1+1$ dimensions can be extended to a rational non-degenerate one (on a bigger 
Hilbert space). To the new 
theory $\2F$ Rehren's analysis \cite{khr2} applies and proves that the category of DHR 
sectors is a modular category in the sense of Turaev. Since the identification of the
degenerate sectors with a group dual \cite{dr3,dr4,dr5,dr1} has a categorical analogue 
\cite{dr6} it is very natural to conjecture that there exists an abstract version 
of this construction for braided $C^*$-tensor categories. We conclude this paper with 
the following
\bconj Given a braided $C^*$-tensor category with dual objects, direct sums and 
subobjects (thus automatically a ribbon category by \cite[Sect.\ 4]{lro}) one can 
construct a non-degenerate category of the same type using methods from \cite{dr6}. 
If the original category is rational, the same is true for the new one which thus is 
modular. \econj
Work on this conjecture is in progress.
\\ \\ \\ \noindent
{\it Acknowledgments.} I am greatly indebted to K.-H.\ Rehren for proposing the subject 
of this work and for many useful discussions. This paper would not have been possible
without his earlier work in \cite{khr2,khr1}. I thank R.\ Longo and J.\ Roberts
for drawing my attention to a serious gap in the first version of this paper. I am
grateful to them as well as to R.\ Conti, D.\ Guido and R.\ Verch for useful discussions.
Again, many thanks are due to P.~Croome for a very thorough proofreading of an earlier
version.

\end{document}